\documentclass[12pt]{iopart}
\usepackage{amsthm}
\usepackage{mathptmx}
\usepackage{color}
\usepackage[dvips]{graphicx}

\newcommand{\RK}{\texttt{RK8NB}~}
\begin{document}

\title{A New Method to Integrate Newtonian N-Body Dynamics}

\author{V. Parisi \& R. Capuzzo-Dolcetta}

\address{Dep. of Physics, Sapienza, Universit\'a di Roma, p.le A. Moro 2, 00185, Roma, Italy}
\ead{valerio.parisi@uniroma1.it, roberto.capuzzodolcetta@uniroma1.it}
\vspace{10pt}
\begin{indented}
\item[]January 2019
\end{indented}

\begin{abstract}
In this note we approach the classical, Newtonian, gravitational $N$-body problem by mean of a new, original numerical integration method.
After a short summary of the fundamental characteristics of the problem, including a sketch of some of its mathematical and numerical issues, we present the new algorithm, which is applied to a set of sample cases of initial conditions in the `intermediate' $N$ regime ($N=100$).
This choice of $N$ is not due to algorithm limitation but just for computational convenience, in what this preliminary work aims mainly to the presentation of the new method and so we wanted just to provide an acceptable although statistical significant comparison with other integration schemes. 
The proposed algorithm seems to be fast and precise at the same time and so promising for further, more realistic, tests and scientific applications.
\end{abstract}

%
\vspace{2pc}

\section{Introduction}
The role of gravity in Astrophysics is fundamental. In the classical (non relativistic) vision, gravity is considered as an infinite range force which acts instantaneously in every point of the space, causing the motion of massive objects in an absolute time frame. In spite of successes of General Relativity in explaining a series of phenomena not accounted for by Newtonian mechanics, it is clear that, as a matter of fact, Newtonian approach is sufficient to deal with enormous precision with a huge majority of relevant space science and astronomical phenomena.
Actually, the regime of \textit{weak} field is that typical of interplanetary space probes, as well as of artificial satellites around Earth (excluding particular cases as that of GPS constellations, whose precision requires accounting for general relativistic effects). The same for both celestial mechanics case (solar system dynamics) and stellar dynamics (star motion in galaxies and galaxy motion in clusters of galaxies). General Relativity becomes actually an unavoidable tool of investigation just when dealing with dynamics around compact (white dwarves, neutron stars) or super compact objects (black holes), or when dealing with the cosmological, large scale, context. 

All this makes evident how in celestial mechanics and stellar dynamics, the classical Newtonian approach to the study of the motion of a system of gravitating bodies is still of enormous importance and, from pioneering work of Poincar\'e in the XIX century \cite{Poin90}, it became a relevant topic of investigation for both theoretical and applied Physics, Astrophysics and Space Sciences. 

The classical, gravitational, $N$-body problem is highly non-linear and chaotic and has only 10 integrals of motion. Consequently, although a formal solution by series of the generic $N$-body problem has been presented by \cite{Wang91}, enlarging the solution by series found by \cite{Sund13} in the case of $N=3$ and non zero angular momentum, it has no practical application because its convergence requires an overwhelming amount of computations. So, any approach to $N\geq 3$ cases requires an approximate numerical frame to work with.

The numerical integration of the $N$-body motion has unavoidably to face with two intrinsic difficulties: the \textit{ultraviolet} ($UV$) and \textit{infrared} ($IR$) divergence of the bodies' Newtonian interaction potential.
$UV$ divergence is that of the mutual force at vanishing inter-body distance, while $IR$ divergence corresponds to that Newtonian force never vanishes. These two problems make the search for a good solution of the $N$-body problem a very stiff problem. On a side, $UV$ divergence requires a very short time stepping to properly deal with close encounters. These short time steps make, unavoidably, the integration to get stuck because of the $O(N^2)$ computational complexity as due to the need ($IR$ divergence) of accounting of all the pairs in the system. Consequently, specific algorithms for both force evaluation and time integration have been developed in the past.
Usually, the best compromises are thought to deal with ranges of values for $N$: i) \textit{few} body problems $N < 10$; ii) \textit{small} $N$-body problems ($N\leq 100$; iii) \textit{intermediate} $N$-body problems $N\leq 10^6$; iv) \textit{large} $N$-body problems $N>10^6$. Each of these classes corresponds to astrophysical interesting situations, from the realm of celestial mechanics (typical few body systems) to that of galaxy dynamics and large scale structures in the Universe.

In this paper we present a new method to approximate solutions of the Newtonian $N$-body problem and we give some tests and comparison with some other, classical, numerical approximations.
Specifically, we compare some of the outputs of our method with a low order, but symplectic,  algorithm (Euler- Cromer) and with a high order, non symplectic, one ($8th$ order Runge-Kutta).

\section{The classical, Newtonian, N-body problem}
In an inertial coordinate frame, the equations of motion of $N$ point of masses interacting via Newtonian gravity and with given initial conditions are (for $k=1,2,...,N$)

\begin{equation}
\label{eq:1}
\left\{
\begin{array}{lll}
\ddot{\textbf{r}}_{k}=G\sum_{j=1\atop j\neq k}^{N}
{m_{j}}\frac{\textbf{r}_{j}-\textbf{r}_{k}}{r_{jk}^{3}}\\
\textbf{r}_{k}(0) =\textbf{r}_{k0}\\
\dot{\textbf{r}}_{k}(0) = \dot{\textbf{r}}_{k0},
\end{array}
\right.
\end{equation}

where the dots denote time derivatives, $G$ is the gravity constant, $m_{j}$ is the mass of the $jth$ particle, 
$\textbf{r}_{j}$ is its position vector, and $r_{jk}\equiv |\textbf{r}_{j}-\textbf{r}_{k}|$ 
is the Euclidean distance between particles $j$ and $k$. 
The set
of equations (\ref{eq:1}) corresponds to a conservative, reversible,
dynamical system, with $3N$ spatial degrees of freedom and $6N$ in the phase space. The right
hand side of the generic equation  in the system (\ref{eq:1}) represents
the total force on the $k$-th object as due by all the other $N-1$ point
masses.  This force is obtainable
as gradient (done respect to the particle coordinates) of the (potential) function 

\begin{equation}
\label{eq:2}
U=\frac{1}{2}\sum_{
(j,k)=1\atop k\neq j}^N
\frac{Gm_{j}m_{k}}{r_{jk}},
\end{equation}
so that the equations of motion, can be written as a system of $6N$ first order differential equations

\begin{equation}
\label{eq:3}
\left\{
\begin{array}{llll}
\dot{\textbf{r}}_{k}=\textbf{v}_{k} \\
\dot{\textbf{v}}_{k}=\frac{1}{m_{k}}{\nabla}_{k}U \\
\textbf{r}_{k}(0) =\textbf{r}_{k0}\\
\textbf{v}_{k}(0) =\textbf{v}_{k0},\\
\end{array}
\right.
\end{equation}

where $\nabla_{k}$ is the gradient operator acting over the coordinates $x_k,y_k,z_k$ and $\dot{\textbf{v}}_{k}$ is the acceleration of the $k$-th particle.
From the point of view of the theory of systems of differential equations,
the system (\ref{eq:3}) is a system of $6N$ first-order equations
in the $6N$ variables ($x_1,y_1,z_1;v_{x1},v_{y1},v_{z1}; ...;x_N,y_N,z_N;v_{xN},v_{yN},v_{zN}$).

Once that the equations of motions are written in the form (\ref{eq:3}), their numerical solution can exploit of a plethora of  available numerical algorithms. Anyway, only a very careful numerical treatment allows to obtain a reliable solution, overcoming the $UV$ and $IR$ divergence of the interaction force and the strong non-linearity of the right hand side of the system.
At this regard we limit to cite the book \cite{Aarbook},  which is comprehensive of most of the aspects of the hard topic of gravitational $N$-body simulations.

\section{The algorithm}
As well know, the explicit solution of the classic gravitational $N$-body problem, does not exist,  therefore it is necessary to resort to numerical approximations, as we said in the above section.
 
The gravitational, Newtonian, $N-$ body problem is chaotic, anyway if dealing with few bodies and restricting to a time  "small" compared to the system's Lyapunov time, it can be considered as almost "deterministic". Therefore, the solution can be approached {\it ad libitum} by mean of standard integration algorithms suited to systems of ordinary differential equations, both symplectic and not. 
On the other side, when aiming to the study of a many-body motion over a time which is "long" compared to Lyapunov time, what it can be actually looked for is no more an approximation of the "exact" solution,  but, rather, only the best possible estimate of the statistical properties of the possible solutions.
 
In such cases, and which such aims, the defect of classical integration algorithms is that of being unnecessarily {\it precise}, expensive, delicate, and complex. Consequently, we are here proposing a totally different type of algorithm,  which, in spite of a loss of (unnecessary) precision, results cheaper, simpler and more robust in giving reliable evaluation of relevant statistical indicators of the $N-$ body system behaviour
 
 We give here only a short summary of the idea behind our scheme (which we will hereafter referred to as \textit{bizarre} method), whose extensive description and accurate testing will be given in a specific paper (in preparation).
 
The key idea of this type of algorithm is that of replacing the gravitational potential  between two bodies, smoothly varying with the inverse of their distance, with a step-wise potential, kept constant along a certain number of steps. In such a scheme, the gravitational attraction as function of the mutual distance assumes a more bizarre behaviour (which justifies our naming): for almost all distances it is absent except for some distances for which it is infinite.
In this scheme, the system of $N$ bodies makes $N(N-1)/2$ pairs. 
In our algorithm every object ($j-th$ particle) is thought moving freely in space until it reaches the first potential energy step given by one ($k-th$ particle) of the other $N-1$ bodies. 
Of course each object in the generic pair (masses $m_j$ and $m_k$) can be treated as one single body of mass $m_j+m_k$ which is moving respect to their center of mass with a velocity $\Delta v$. During their travel, the two objects 
conserve that velocity until they cross an energetic level step (see Fig. 1). At crossing, three possibilities arise: i) \textit{slowdown in distancing}, ii) \textit{acceleration in approach}, iii) \textit{reflection}.
At each of the three cases corresponds a specific updating of the velocity respect to the center of mass, $\Delta v$.
Accordingly, the velocity of the two particles is updated according to the classical expressions

\begin{equation}
\label{eq:4}
\left\{
\begin{array}{ll}
v_j &=v_{cm}+\frac{m_k}{m_j+m_k}\Delta v, \\
v_k &=v_{cm}-\frac{m_j}{m_j+m_k}\Delta v. \\
\end{array}
\right.
\end{equation}

where $v_{cm}$ is the center of mass velocity.

The dynamics of bodies subjected to this potential admits solutions that resemble the "right" solution to arbitrary precision provided an improvement of the approximation of potential, which passes through an increasing refinement of the stepping. Our standard step size procedure is that of a definition of the radial size of the step as obtained by a geometric progression in a given $[r_{min},r_{max}]$ interval. This choice has been proven to be a good compromise between the search for high resolution and computational convenience. 
 
The remarkable property of this approach is that it makes possible to calculate analytically and numerically the "right" solution of the approximate problem, which is fully conservative, so that the  energy and momentum conservation is guaranteed up to machine precision.  From the preliminary tests made, the method seems to be particularly promising for the study of the dynamics of stellar clusters of intermediate size.

\begin{figure}
\label{fig:1}
\begin{center}
\includegraphics[width=10cm,height=8cm]{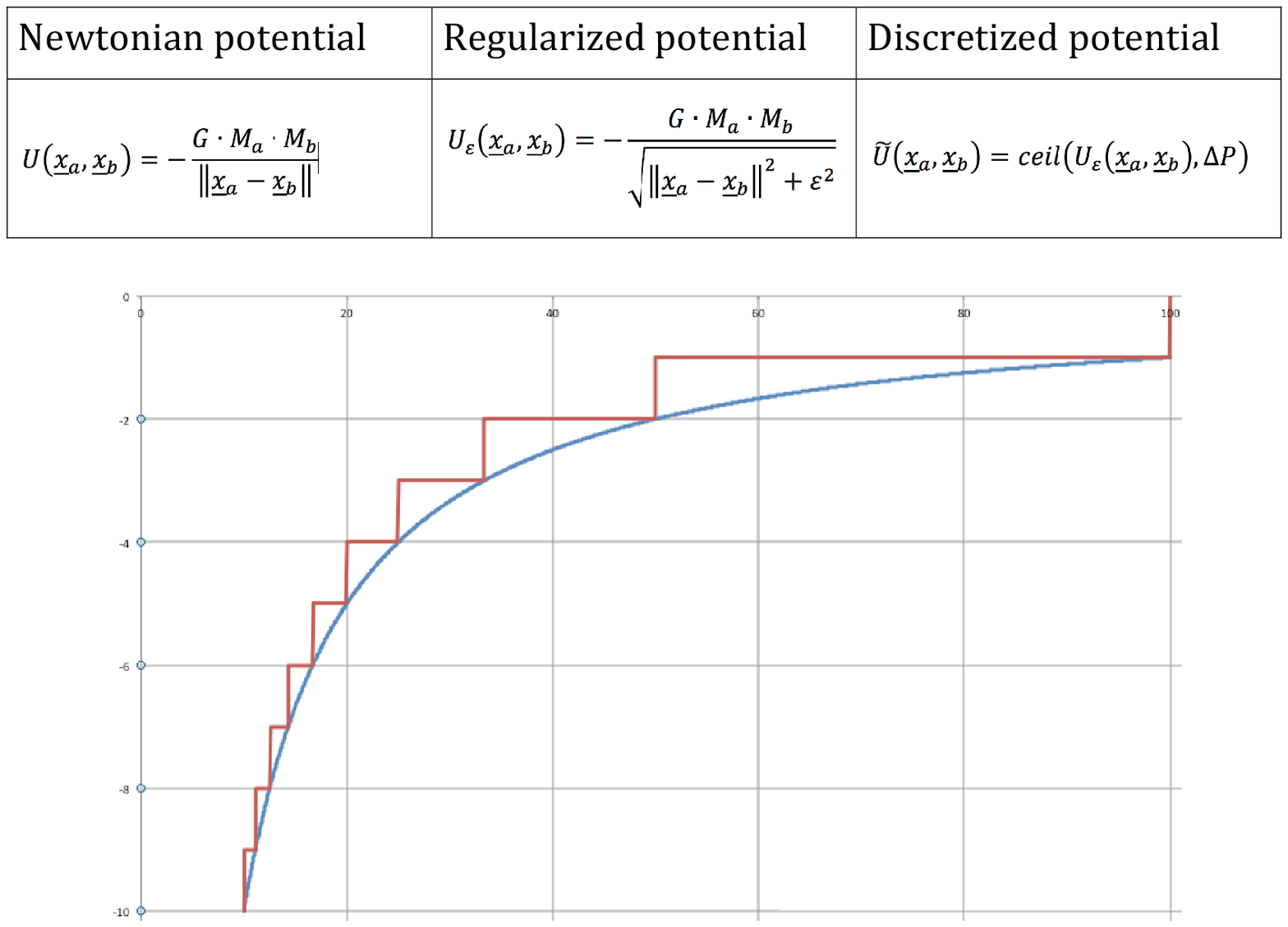}
\end{center}
\caption{In red we show the step-wise approximation of the body interaction potential
(note in the caption on top we use capital m for the mass value and underlined $x$ for position vectors).}
\end{figure}

\section{Preliminary results}
To test the validity, both in terms of precision and speed, of our algorithm we decided to perform a set of numerical simulations with the aim of providing sufficient statistical reliability.
For computational convenience, we smoothed (as it is commonly done \cite{Aarbook}) the pure Newtonian interaction potential by mean of the introduction of a softening parameter, $\epsilon$, such that the ($k,j$) pair potential, $U_{kj}$, becomes

\begin{equation}
 U_{kj}=G \frac{m_km_j}{\sqrt{r_{kj}^2+\epsilon^2}}.   
\end{equation}
Of course, in this approximation the mutual force vanishes, instead of diverging, for vanishing $r_{kj}$., avoiding $UV$ divergence. So, to keep a good approximation of the overall Newtonian behaviour of the simulated system, $\epsilon$ must be chosen as a fraction of the closest neighbhour inter-particle distance, $\langle d_{cn}\rangle =\alpha L N^{-1/3}$ where $L$ is the size of the system and $\alpha$ is a \textit{safety} coefficient, which we set to $0.05$.

We exploited time integrations of the motion of $N=100$ equal mass particles, assuming units such that the total mass, $M$, is the mass unit, and the radius of the sphere containing initially the $N$ bodies, $L$, is the length unit. The further assumption that the gravitational constant is set to $G=1$ leads to $T=L^{3/2}/\sqrt{GM}$ as time unit (this time is referred to as \textit{dynamical} or \textit{crossing} time). This choice of units allows a full adimensionalization of the motion equations, so that a simple rescaling \textit{a posteriori} of results gives positions, velocities, energy, etc. in physical units, once a specific assumption for initial values of $L$ and $M$ are done.

The initial conditions for the $N=100$ bodies have been randomly sampled from a uniform distribution in both position and velocity spaces. Actually, after uniform velocity sampling, we re-scaled the obtained velocities in order to have a chosen value for the so-called "virial" ratio, defined as

\begin{equation}
\label{eq:5}
Q=\frac{2K}{|\Omega|},
\end{equation}

where $K$ and $|\Omega|$ are, respectively, the total kinetic and potential energies of the system.
A system with $Q=1$ is called \textit{virialized}. This corresponds to a second time derivative of its polar moment of inertia 

\begin{equation}
\label{eq:6}
\ddot{I}=\frac{d^2I}{dt^2}=\frac{d^2}{dt^2}\sum_{k=1}^N \, m_k \textbf{r}_k\cdot \textbf{r}_k=0,
\end{equation}

i.e. the system is globally stationary.

Given this, we sampled a set of $100$ initial conditions by means of $100$ different initial random \textit{seeds}, for a system with $Q=1/2$, that means a \textit{sub virial} system, which undergoes an initial collapse in a time of the order of $T$ with a subsequent rebound and a further series of damped oscillations around an almost stationary configuration. 

All $N$-body Newtonian systems are expected to virialize in a time of the order of few dynamical times. To virialization follows a secular evolution, which is mainly due to 2-body interactions, on a time scale (relaxation time scale) which is of the order of $(N/\ln N)T$.

\begin{figure}
\label{fig:2}
\begin{center}
\includegraphics[width=7cm,height=6cm]{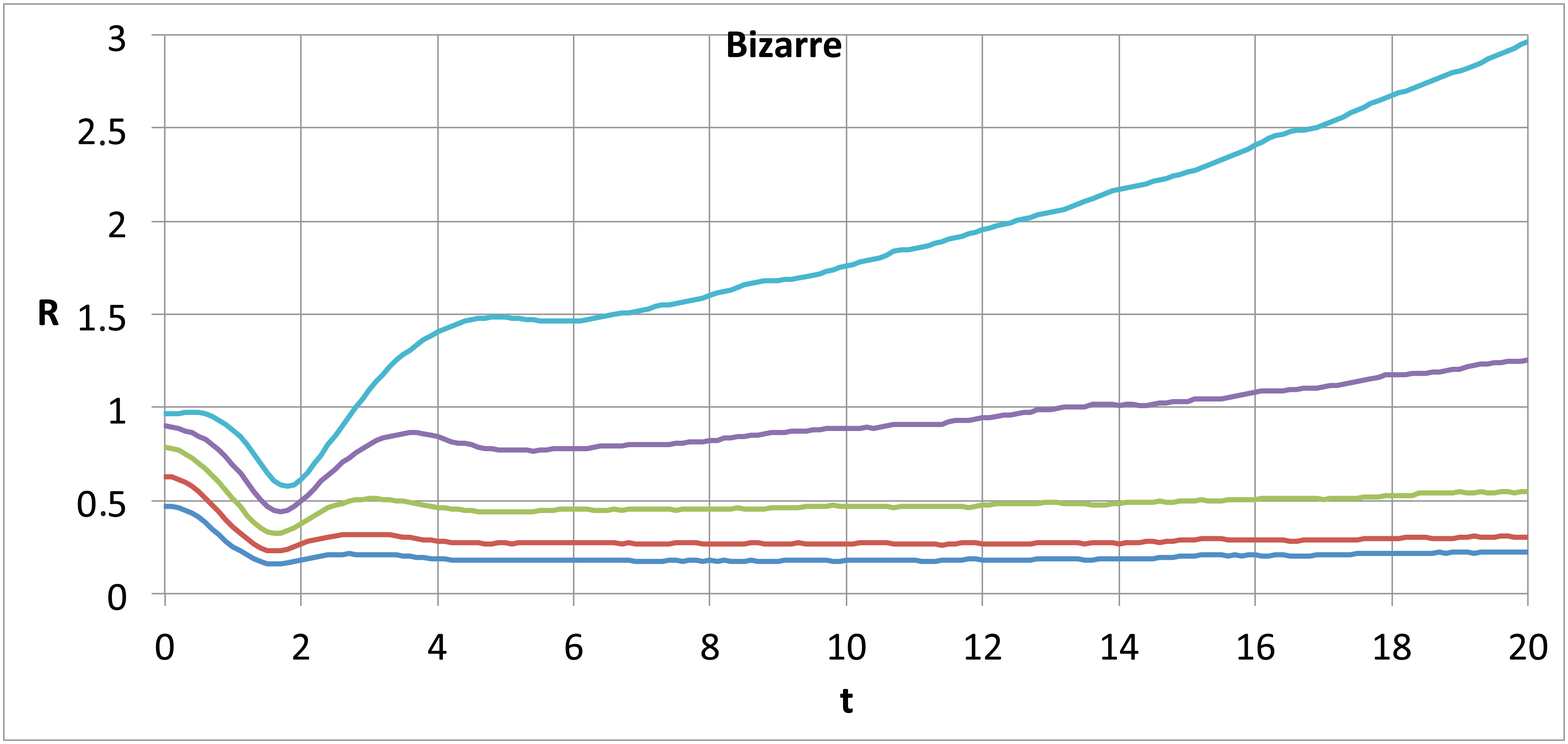}\hfill
\includegraphics[width=7cm,height=6cm]{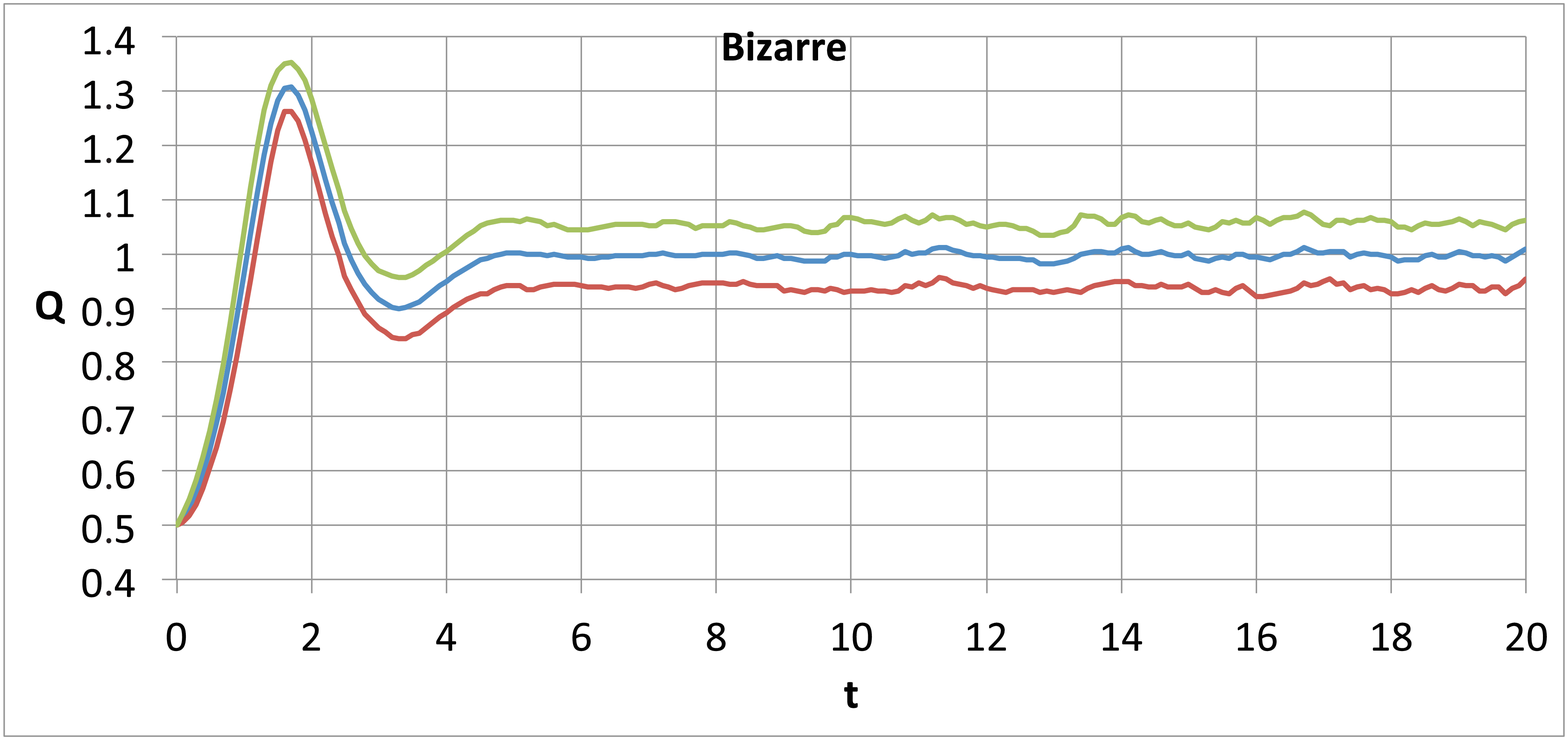}
\end{center}
\caption{Bizarre method. Left panel: the behaviour of Lagrangian radii versus time. Different colours are for different values of mass fraction contained in the Lagrangian radius. Top down: $90\%$, $75\%$, $50\%$, $25\%$, $10\%$.
Right panel: virial ratio versus time. The 3 curves refer to the average over the $100$ simulations and (above and below) the curves corresponding to $\pm 1 \sigma$ curves.}
\end{figure}

\begin{figure}
\label{fig:3}
\begin{center}
\includegraphics[width=7cm,height=6cm]{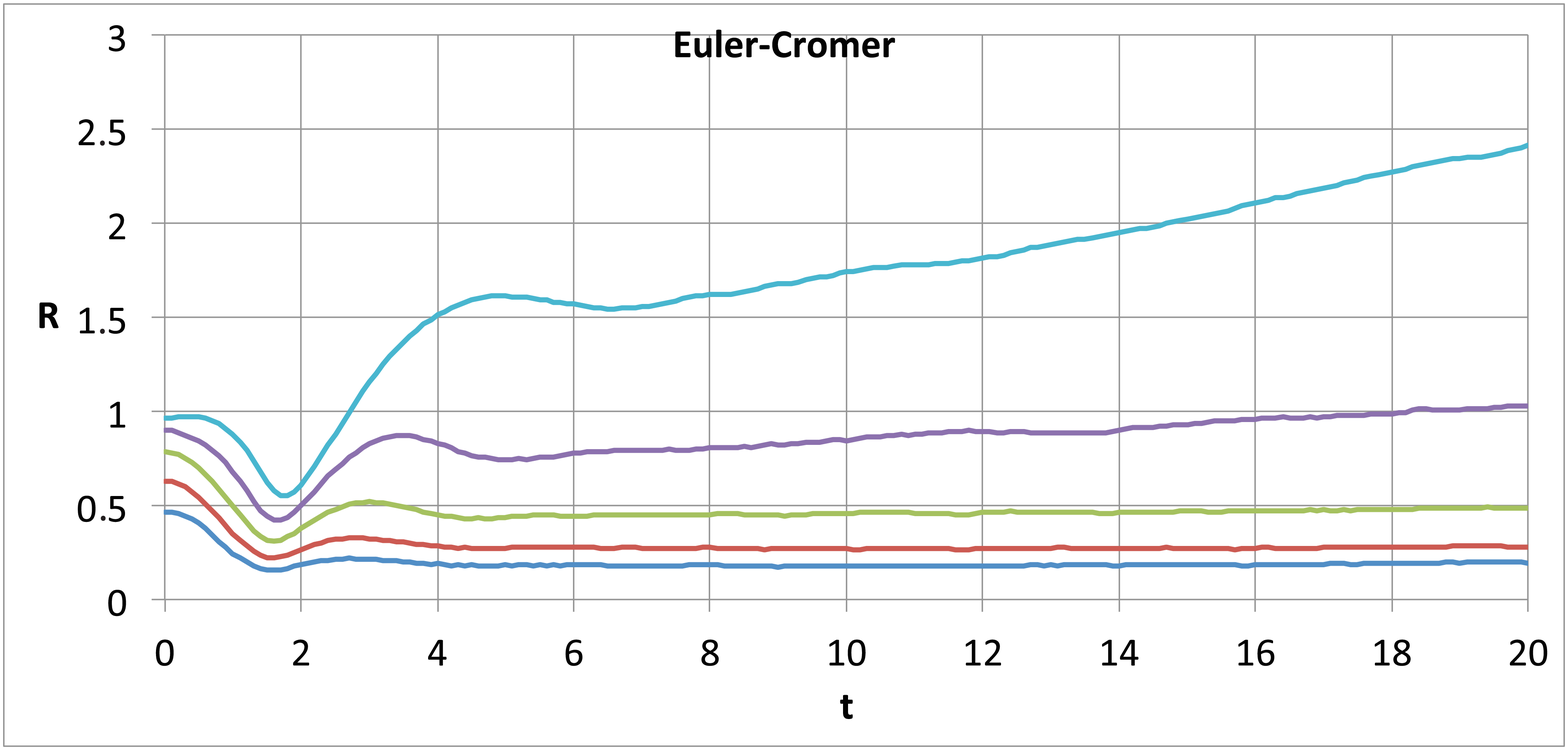}\hfill
\includegraphics[width=7cm,height=6cm]{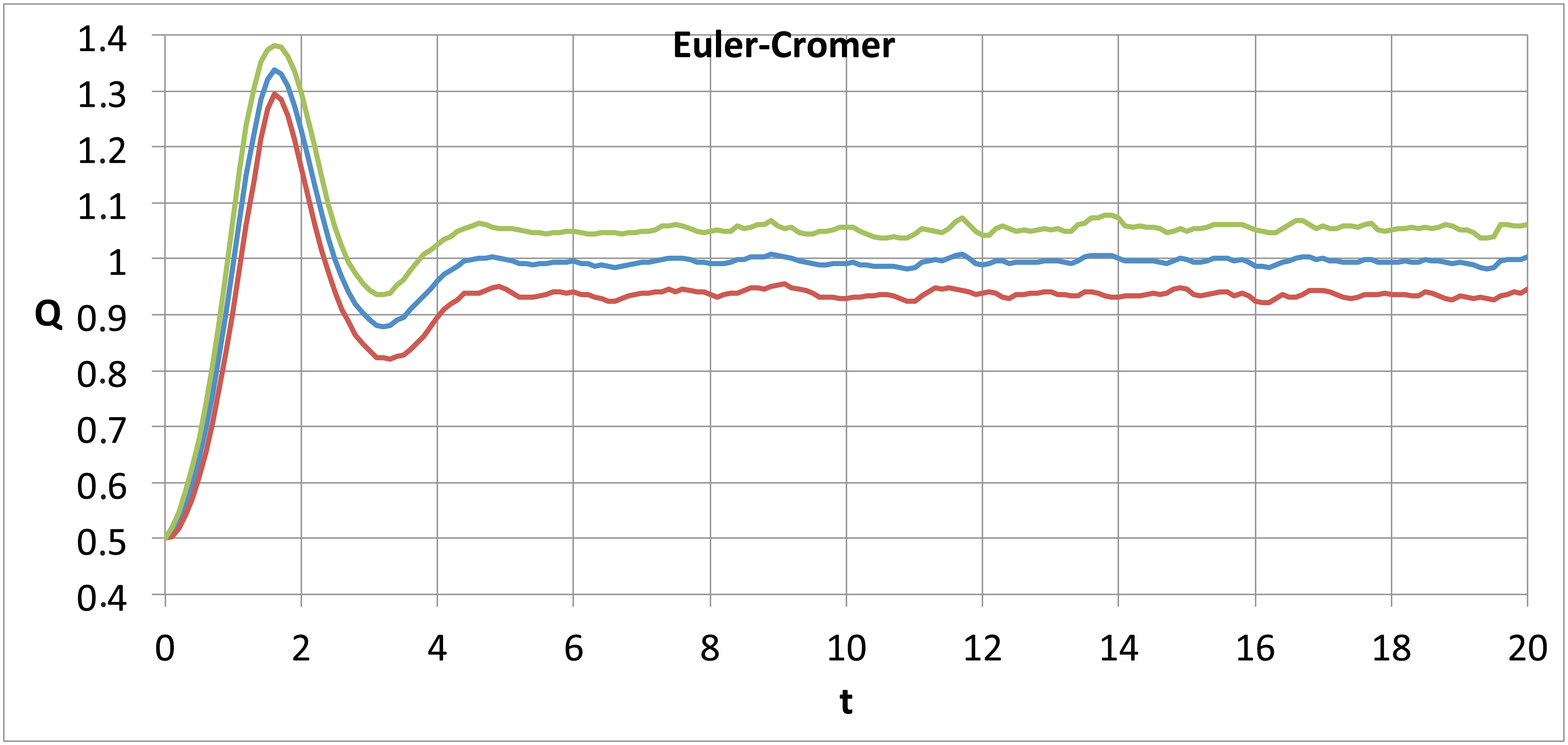}
\end{center}
\caption{Euler-Cromer method. Left panel: the behaviour of Lagrangian radii versus time. Different colours are for different values of mass fraction contained in the Lagrangian radius. Top down: $90\%$, $75\%$, $50\%$, $25\%$, $10\%$.
Right panel: virial ratio versus time. The 3 curves refer to the average over the $100$ simulations and (above and below) the curves corresponding to $\pm 1 \sigma$ curves.}
\end{figure}

\begin{figure}
\label{fig:4}
\begin{center}
\includegraphics[width=7cm,height=6cm]{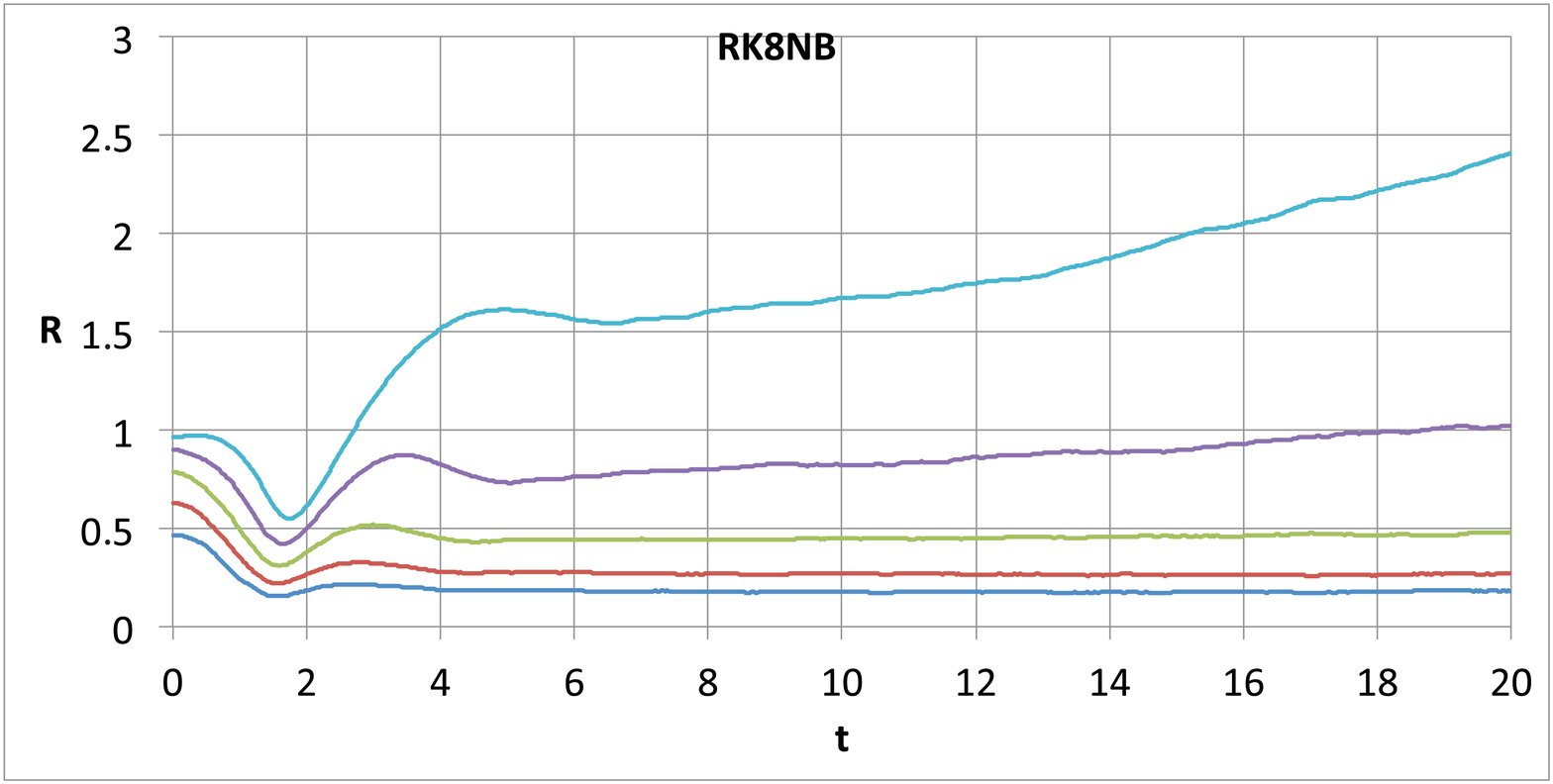}\hfill
\includegraphics[width=7cm,height=6cm]{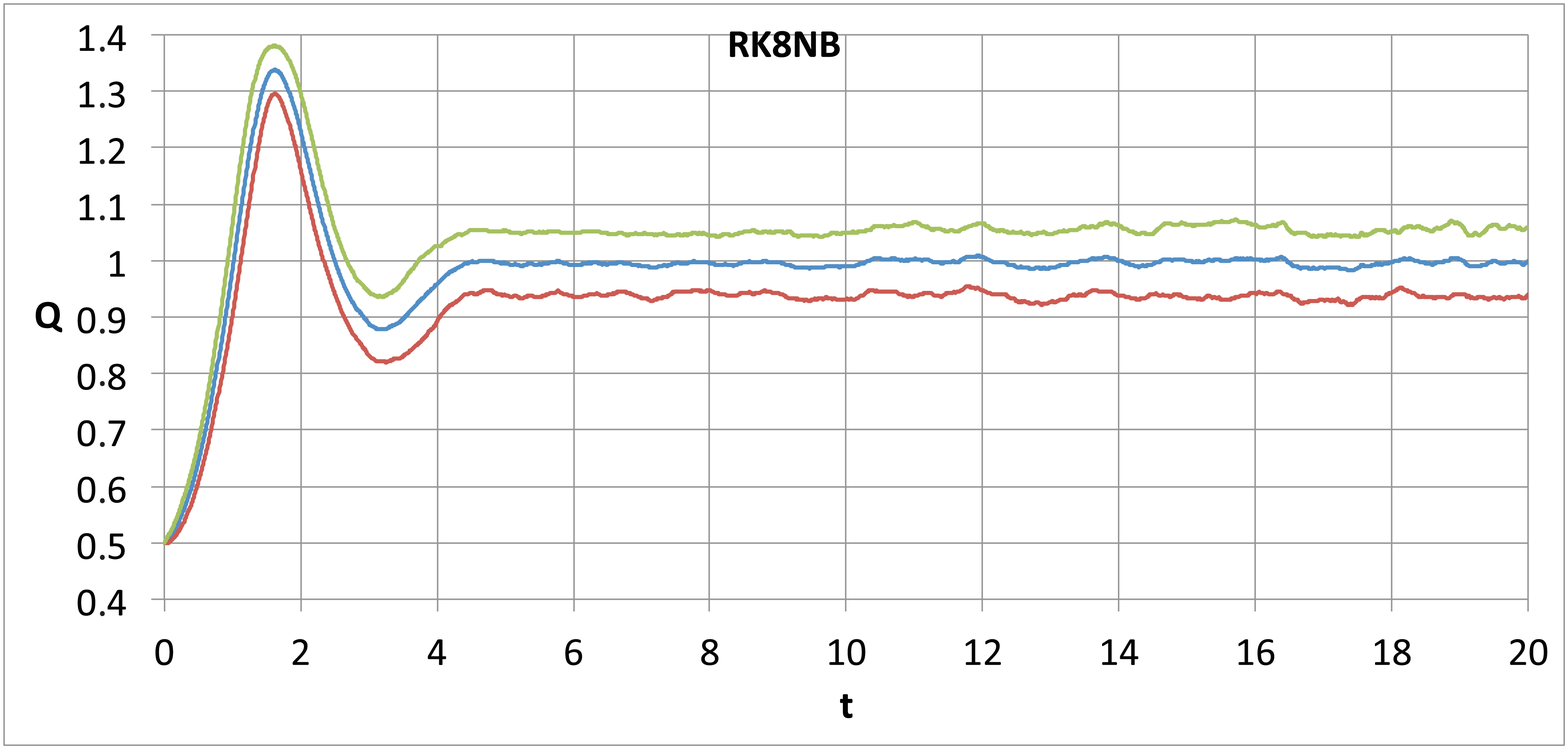}
\end{center}
\caption{Runge-Kutta $8th$ order method. Left panel: the behaviour of Lagrangian radii versus time. Different colours are for different values of mass fraction contained in the Lagrangian radius. Top down: $90\%$, $75\%$, $50\%$, $25\%$, $10\%$.
Right panel: virial ratio versus time. The 3 curves refer to the average over the $100$ simulations and (above and below) the curves corresponding to $\pm 1 \sigma$ curves.}
\end{figure}

As we said, given the chaotic behaviour of classical, gravitational, Newtonian $N$-body systems, significant results might be obtained just when considering some statistical properties, at least when a long term evolution is analyzed. 

For the sake of this preliminary work we limited our attention to some global parameters like, indeed, the time evolution of 
\begin{itemize}
\item the virial ratio, Q;
\item the Lagrangian radii (which are the radii of the spheres containing a given fraction $\beta$ of the total mass of the system), $R_\beta$;
\item the number, $N_e$, of escapers, i.e. bodies which become unbound to the system after energy exchange with other objects of the system.
\end{itemize}

Due to that the individual body mechanical energy is not conserved, the criterion to define the $jth$ body as true escaper cannot be based just on $E_j>0$, because the positive energy could be reverted again to negative due to other individual or cumulative interactions. So, a safer criterion, that we used, is that to state that the $jth$ object becomes unbound to the system when it gets $E_j>0$
and $r_j>2 R_{1/2}$ ($r_j$ is the distance to the system center of mass and $R_{1/2}$ is the $50\%$ Lagrangian radius, also called half-mass radius).

\subsection{Simulation comparisons}
To check properties of our new algorithm on both sides of precision and computational efficiency, we compared some of its results with those obtained with two other well known integrators, based, respectively, on our implementation of the simplest symplectic integrator (order one), i.e. the semi-implicit-Euler method (also called Euler-Cromer, see \cite{Cro81}, although it has been developed first in \cite{Par77}) and of an implementation (called \RK, \cite{Cri19}) of a high order ($8th$ order), fully explicit, Runge-Kutta method (see, e.g. \cite{But08}). 

The comparison is just preliminary, and does not pretend to be exhaustive. As a matter of fact, we obtained results at very different levels of accuracy: with \RK the energy is conserved with a fractional error always less than $10^{-9}$ and angular momentum is conserved within $10^{-12}$, while with the Euler-Cromer we were at $10^{-6}$ level. To keep this high level of accuracy, \RK fruits of a proper adaptive time stepping technique. 

In Fig. 2 we show, for our new algorithm, the behaviours of the Lagrangian radii and virial ratio as functions of time, in average and variance computed over the $100$ representations we performed.
As expected, given the sub-virial initial conditions, innermost Lagrangian radii show a decrease and a stabilization while outer radii after an initial contraction show a rebound which is consequence of the acceleration gained by some of the objects in the system during the high density phase around one characteristic time. The initial violent dynamics and trend to virialization is clearly shown by the $Q(t)$ plot in the right panel of Fig. 2. A peak of $Q\simeq 1.2$ is reached at $t\simeq 1.8$ ,
followed by a system re-expansion and a virialization ($Q \rightarrow 1$).

These behaviours are very similar to those obtained from the other two methods used, Euler-Cromer and $8th$ order Runge-Kutta. For a graphical comparison, we report in Fig. 3 and Fig. 4 results for the Euler-Cromer and Runge-Kutta $8th$ order methods, respectively.
Let us note the high similarity, but for some discrepancy in the behaviour of the $90\%$ Lagrangian radius for the bizarre method respect the other 2 well established methods and codes.
The discrepancy is that of a higher expansion rate of the $90\%$ radius that, in a system of $100$ particles, is very sensitive to fluctuations because of the (small) number of escapers. It suffices that by numerical error few of the escapers gain ane even moderate excess of kinetic energy to induce such behaviour and discrepancy. We are convinced that this can be easily cured by a better implementation of this new method, whose present version was not deeply analyzed and improved because of our willing of presenting it as new and, possibly, alternative method to deal with gravitational $N$-body systems.

With all the methods used, the number of escapers is low (below $7\%$) so to make statistically insignificant any specific consideration.

On the computational side, the bizarre algorithm proved to be fast, at least in the test simulations presented here. In the average, the integration up to $20$ crossing times required 
14 sec on a early 2013 Mac book Pro equipped with an Intel I7 2.7 Ghz processor and a 1600 MHz DDR3 RAM. Although at the light of the above mentioned limitation on reliability of the present computational speed evaluation, this means $\sim 60$ times faster than our Euler-Cromer implementation and $\sim 8$ times faster of $8th$ order  Runge-Kutta. 

\section{Conclusions}
We presented a new method, which we called \textit{bizarre}, to follow the time evolution of a system of $N$ point masses interacting via classical Newtonian gravitational force. 

The method has the characteristic to be fully conservative because all the time updates are done by pair, imposing both energy and momentum conservation.
The aim of this preliminary research note was mainly the presentation of this method and a check of how much this method is able to reproduce expected general statistical behaviours of moderate $N$ ($N=100$) equal mass bodies. A comparison with a well established symplectic method is fully positive because both the Lagrangian radii time evolution and that of the virial ratio are very similar. On a performance side, without any ambition to state clearly about it, we note that the speed of our present implementation of the bizarre method is $60$ times that of the Euler-Cromer.
and $8$ times that of a highly performant $8th$ Runge-Kutta code, at least in the cases studied here.

Future, logical, developments of this work will be a careful optimization, in order to investigate about its real potential in the study of $N$-body simulation of actual astronomical significance.

\ack
We acknowledge the help given by Matteo Cristiano in running the comparison tests with the \RK code.

\vskip 0.25truecm

\bibliographystyle{unsrt}
\bibliography{ASetal2015}

\vfill\eject

\end{document}